\documentclass[aps,prd,superscriptaddress,showpacs,nofootinbib,eqsecnum]{revtex4}

\input{epsf}
\usepackage{latexsym}

\begin{document}

\title{Noise and Dissipation During Preheating}

\author{Sergio E. Jor\'as}
\email{joras@if.ufrj.br}
\affiliation{Instituto de F\'{\i}sica, Universidade Federal do Rio de
Janeiro, Caixa Postal 68528,  21945-970 Rio de Janeiro, RJ, Brazil}

\author{Rudnei O. Ramos}
\email{rudnei@dft.if.uerj.br}
\affiliation{Departamento de F\'{\i}sica Te\'orica,
Universidade do Estado do Rio de Janeiro,
20550-013 Rio de Janeiro, RJ, Brazil}

\begin{abstract}

We study the consequences of noise and dissipation for parametric
resonance during preheating. The effective equations of motion for
the inflaton and the radiation field are obtained and shown to
present self-consistent noise and dissipation terms. The equations
exhibit  the usual parametric resonance phenomenon, allowing for
exponential amplification of the radiation modes inside the
instability bands. By focusing on the dimension of the border of
those bands we explicitly show that they are fractal, indicating the
strong dependence of the outcome in the initial conditions. The
simultaneous effect of noise  and dissipation to the fractality of
the borders are then examined.

\end{abstract}

\pacs{98.80.Cq}

\maketitle

\section{Introduction}

In standard inflationary scenarios of the early universe, the
transition
from the inflationary regime itself to the radiation dominated phase
(or standard big bang regime) is accomplished through a short period
of reheating after inflation, when the inflaton field, by rapidly
oscillating at the bottom of its potential in a time scale shorter
than the Hubble time, releases all of its energy in the form of
light particles that it may be coupled to. Recent studies on this
reheating phase have shown the possibility of emergence of new and
interesting phenomena, like explosive particle production,
nonequilibrium  symmetry restoration, among others effects
\cite{reheat}. These effects are directly related to the possibility
of the oscillating inflaton field  to drive the fluctuations of the
coupled fields to regimes of parametric amplification, or
parametric resonance, where particles are  efficiently produced.

The parametric resonance, or preheating phase, has then been shown to
be a fundamental effect that must take place before the actual
reheating  of the universe through the thermalization of the produced
particles \cite{preheat}. Preheating also can be easily obtained in
very  different models by means of a linear or a quadratic coupling
between the  inflaton and the radiation fields. In all those cases
the inflaton,  oscillating at the bottom of its potential (after the
inflationary phase), behaves  as a periodic mass for the other scalar
fields that are coupled to it. As  a consequence, the equations for
the modes $k$ of the fluctuation  fields present exponential
solutions

\begin{equation}
\chi_k(t)\sim \exp [\pm \mu \cdot t]\;,
\end{equation}
where $\mu$, the Floquet exponent, may be either real or complex,
depending on the parameters adopted. One is thus able to define the
 so called instability bands in the parameter space, within which the
Floquet exponent is real, indicating exponentially (de)increasing
solutions. The regions of the parameter space where $\mu$ is
imaginary, on the other hand, correspond to oscillatory solutions
and are thus called stability bands.

One important aspect concerning the band structure and
exponential growth of fluctuations is the possible chaotic behavior
associated to the very existence of the bands, both at their borders
and inside them. Our concern in this paper is thus to determine the
dimension of the borders between two such bands. We will also be
interested in calculating the dimension of the border between two
different values  of the Floquet exponent inside a given instability
band. As we will  show, all of them are fractal, unambiguously
characterizing them as chaotic ones. As such, they prevent one from
telling if a given mode, assumed close enough to a border, will be
amplified or not. Thus, the outcome  of the preheating phase will
strongly depend on the state of the  universe right after the
inflationary period.

The fractality, or chaoticity, of the borders separating instability
bands is, in a sense, not a completely unexpected result. For
example, in a recent work \cite{joras} the authors argued for the
close relation between the presence of chaos and parametric
resonance. In fact, the equation of motion for the fluctuation fields
coupled to the inflaton field, viewed as a classical equation of
motion, resembles that of an harmonic oscillator driven by a periodic
external force, which is  well known to exhibit chaotic behavior
under appropriate conditions. A few other papers have also studied
chaos in the context of preheating, as  by the authors of Ref.
\cite{felder}, although they have studied chaos  {\it after} the
preheating period. In general, the dynamics of interacting fields may
show the possibility of chaotic dynamics, as has been  shown
recently in Refs. \cite{RF,ROR}.

One important aspect associated to the microscopic physics of
interacting fields raised in Refs. \cite{RF,ROR} was the importance
of dissipation in the dynamics of coupled system of fields and the
influence of dissipation on the degree of chaoticity of the field
dynamics. Recall that typically the effect of dissipation is to damp
the fluctuations on the system and consequently tends to suppress
possible chaotic motions and makes field trajectories in phase space
to tend faster to the system asymptotic states. This is indeed the
case for  the dynamics of background (or classical) field
configurations. However,  the fluctuation (or quantum) modes of the
fields that experience  parametric resonance are not only subject to
dissipation but also to stochastic noise, being both related by a
generalized fluctuation-dissipation relation \cite{GR}.

Previous works have also studied the influence of noise during
preheating. For example the authors of Refs. \cite{noise1,noise2}
have studied the consequences to preheating when a somewhat {\it ad
hoc} noise is added as a mass term to the secondary field's equation
of motion. Here, in contrast, we will take the equation for the
inflaton  as a typical ensemble averaged equation of motion (as we
should expect  for a background field configuration) and noise is
only included in the effective equation of motion for the fluctuation
field coupled to the inflaton. This equation is self-consistently
derived for both dissipation and noise through the usual method of
Schwinger's closed time path formalism of real time (for previous
references, see for instance Refs. \cite{GR,hu1,morikawa}). The
simultaneous effect of  noise and dissipation for the reheating have
been also considered recently  by the authors of Ref. \cite{hu2}
where they also derived  self-consistently the effective equations of
motions for both the inflaton and the radiation fields. Their main
interest was, however, in the final  stages of reheating, not
including therefore the consequences of both noise  and
dissipation for the preheating phase. This is our main concern in
this paper, which we will specialize in the simultaneous effect of
noise  and dissipation on the fractality of the borders of the
instability bands and how these microscopic physics associated to the
dynamics of the fields may alter the parametric resonance phenomenon
during  preheating.

The chaotic behavior associated to the parametric resonance stage is
then studied by means of the determination of the fractal dimensions
of the borders of the instability bands. As argued before in Refs.
\cite{RF,ROR} (see also \cite{levin}), the fractal dimension gives a
topological measure of chaos for different space-time settings and it
is a quantity invariant under coordinate transformations, yielding
then  an unambiguous signal for chaos in cosmology and general
relativity problems in general. The dimension of the border provides
thus a  robust characterization for chaos as opposed to Lyapunov
exponents which may change sign by a simple change of observer
\cite{george}.

This paper is organized as follows: in Sec. II we consider a simple
model leading to preheating consisting of an inflaton field $\phi$ of
chaotic\footnote{Here, ``chaotic inflation'' only means we use a
single-well potential for the inflaton, and has no relation
whatsoever to the actual chaotic behavior we describe later on.}
inflation with quadratic potential coupled to a set of radiation
scalar fields  $\chi_j$ that $\phi$ can decay to. Additionally we
allow the radiation fields  to decay to fermion fields $\psi_k$ (in
this model the final radiation energy density is composed of fermion
matter fields). The effective equations of motion for both the
inflaton $\phi$ field and the  radiation $\chi_j$ fields are then
explicitly derived. In Sec. III we present  our
numerical studies concerning the parametric resonance in our model
and the computation of the fractal dimension of the border of the
instability bands. {}Finally in Sec. IV we give our concluding
remarks.

\section{Deriving the Effective Equation of Motions}

We will study here the simplest model of an inflaton field $\phi$
coupled to a set of scalar fields $\chi_j$, $j=1,\ldots, N_\chi$,
through a trilinear coupling. Although quadratic couplings between
the fields are more common in the study of resonance during
preheating, models with trilinear couplings exhibit the same
phenomenon as well  and had been used before as a toy model in
different analysis  \cite{brand}. We choose this simplest model here
for convenience since it will  allow a more straightforward analysis
of the problem without having to  compute higher order terms, as we
are going to see below. Additionally, we couple to the radiation
scalar fields $\chi_j$ a set of fermion  fields $\psi_k$, $k=1,
\ldots, N_\psi$, thus allowing $\chi_j$ to decay into the fermion
fields. These kind of models as this one we will study  here,
allowing cascade decay sequences as $\phi \to 2 \chi_j$ and $\chi_j
\to \psi_k +\bar{\psi}_k$ have shown to exhibit interesting
dissipation properties \cite{BR} and would then be particularly
interesting to  study them in the context of reheating and parametric
resonance after inflation. The model we study here is then described
by the following Lagrangian density

\begin{eqnarray}
{\cal L}  =  \frac{1}{2} (\partial_\mu\phi)^2  +
\frac{m_\phi^2}{2}\phi^2
+ \sum_{j=1}^{N_{\chi}} \left[ \frac{1}{2} (\partial_\mu\chi_j)^2  +
\frac{m_{\chi_j}^2}{2}\chi_j^2
+ \frac{g_{j}^2}{2}
\phi \chi_{j}^2
\right]
+ \sum_{k=1}^{N_{\psi}}
\bar{\psi}_{k} \left[i \not\!\partial - m_{\psi_k}
- \sum_{j=1}^{N_\chi} h_{kj} \chi_j \right] \psi_{k} \;.
\label{model}
\end{eqnarray}

\subsection{The effective equation of motion for the inflaton field}

We can easily obtain the exact equation of motion for the inflaton
field from the model Eq. (\ref{model}). Decomposing the $\phi$ field
into a classical background component $\varphi$ (the classical
inflaton  field) and a quantum fluctuation part as $\phi = \varphi +
\delta \phi$,  with $\langle \phi \rangle=\varphi$ and $\langle
\delta \phi \rangle =0$, then the equation of motion for $\varphi$
readily follows by imposing $\langle \delta \phi \rangle =0$ at all
orders in perturbation theory (the tadpole method). For a homogeneous
inflaton field this gives

\begin{equation}
\ddot{\varphi}(t) + m_\phi^2 \varphi(t) + \sum_{j=1}^{N_\chi}
\frac{g^2_j}{2} \langle \chi^2_j
\rangle =0 \;.
\label{eom1}
\end{equation}
By perturbatively expanding $\langle \chi^2_j \rangle$ in the
$\varphi$  amplitude, we obtain \cite{BR}
\begin{equation}
\ddot{\varphi}(t) +  m_\phi^2 \varphi(t) +
\sum_{j=1}^{N_\chi} g^2_j \int_{-\infty}^t dt' \; g^2_j \varphi(t')
\int \frac{d^3q}{(2 \pi)^3}
{\rm Im}[G_{\chi_j}^{++}({\bf q}, t-t')]_{t>t'}^2 =0 \;,
\label{eom2}
\end{equation}
where $G_{\chi_j}^{++}({\bf q}, t-t')$ is the real time causal
propagator (in momentum-space) for the $\chi$ field. The expression
for $G_{\chi_j}^{++}({\bf q}, t-t')$ can be found for example in
Refs. \cite{GR,BGR,BR}.

Eq. (\ref{eom2}) can also be directly obtained from the standard
rules of the real time formalism. In the absence of fermions, we can
easily functionally integrate the $\chi_j$ field exactly, since Eq.
(\ref{model}) has only quadratic terms, to obtain

\begin{equation}
\Gamma[\phi] = S[\phi] + \sum_{j=1}^{N_\chi} \frac{i}{2} {\rm tr}
\ln [ \Box + m_{\chi_j}^2 +
g_j^2 \phi] \;,
\label{action1}
\end{equation}
where
\begin{equation}
\frac{i}{2} {\rm Tr} \ln [ \Box + m_{\chi_j}^2 +
g_j^2 \phi]  = - i \ln \int_c D \chi_j \exp\left\{
- \frac{i}{2} \chi_j [ \Box + m_{\chi_j}^2 + g_j^2 \phi] \chi_j
\right\}
\label{int chi}
\end{equation}
and the field is integrated over a time path $c$ that goes from
$-\infty$ to $+\infty$ and then back to $-\infty$. This is the
Schwinger's closed time path formalism.

Eq. (\ref{action1}) is a nonlocal equation in $\phi$. By
perturbatively expanding (\ref{action1}) in the inflaton amplitude
$\phi$ (or in the coupling constant $g$, for $g \ll 1$) we obtain
(see for example Ref. \cite{GR})

\begin{eqnarray}
\lefteqn{\frac{i}{2} {\rm Tr} \ln \left[\Box + m_\chi^2 + g^2 \phi
\right] =
\frac{i}{2} {\rm Tr}_c \sum_{m=1}^{+\infty} \frac{(-1)^{m+1}}{m}
G_\chi^m \left( g^2\phi\right)^m=}
\nonumber \\
& & = \frac{i}{2} \sum_{m=1}^{+\infty} \frac{(-1)^{m+1}}{m}
\left(g^2\right)^m {\rm Tr} \int d^4 x_1 \ldots d^4 x_m
G_\chi^{n_1, l_1}(x_1 - x_2) \left[\phi(x_2)\right]_{l_1, n_2}
G_\chi^{n_2,l_2}(x_2-x_3) \ldots
\nonumber \\
& & \ldots  \left[\phi(x_m)\right]_{l_{m-1},n_m}
G_\chi^{n_m,l_m}(x_m-x_1) \left[\phi(x_1)\right]_{l_m,n_{m+1}}
\: ,
\label{trlnsum}
\end{eqnarray}
where $G_\chi^{n,l} (x-x')$ is the real time $\chi$-field propagator
on the contour $c$, given by ($l,n=+,-$) \cite{chou,weert}

\begin{eqnarray}
&& G_\chi^{++}(x-x') = i \langle T_{+} \chi(x) \chi(x')\rangle
\nonumber \\
& & G_\chi^{--}(x - x') = i \langle T_{-} \chi(x) \chi(x')\rangle
\nonumber \\
& & G_\chi^{+-}(x-x') = i \langle \chi(x') \chi(x)\rangle
\nonumber \\
& & G_\chi^{-+}(x-x') = i \langle \chi(x) \chi(x')\rangle \: ,
\label{twopoint}
\end{eqnarray}
where $T_{+}$ and $T_{-}$ indicate chronological and
anti-chronological ordering, respectively. $G_\chi^{++}$ is the
usual physical (causal) propagator. The other three propagators come
as a consequence of the time contour and are considered as auxiliary
(unphysical) propagators \cite{weert}. The explicit expressions for
$G_\chi^{n,l}(x-x')$ in  terms of its momentum space Fourier
transforms are given by  \cite{chou,rivers}

\begin{equation}
G_\chi(x-x') = i \int \frac{d^3 q}{(2 \pi)^3} e^{i {\bf q} . ({\bf x}
 -
{\bf x}')}
\left(
\begin{array}{ll}
G_\chi^{++}({\bf q}, t- t') & \:\: G_\chi^{+-}({\bf q}, t-t') \\
G_\chi^{-+}({\bf q}, t- t') & \:\: G_\chi^{--}({\bf q}, t-t')
\end{array}
\right) \: ,
\label{Gmatrix}
\end{equation}
where

\begin{eqnarray}
&& G_\chi^{++}({\bf q} , t-t') = G_\chi^{>}({\bf q},t-t')
\theta(t-t') + G_\chi^{<}({\bf q},t-t') \theta(t'-t)
\nonumber \\
& & G_\chi^{--}({\bf q}, t-t') = G_\chi^{>}({\bf q},t-t')
\theta(t'-t) + G_\chi^{<}({\bf q},t-t') \theta(t-t')
\nonumber \\
& & G_\chi^{+-}({\bf q} , t-t') = G_\chi^{<}({\bf q},t-t')
\nonumber \\
& & G_\chi^{-+}({\bf q},t-t') = G_\chi^{>}({\bf q},t-t')
\label{G of q}
\end{eqnarray}
and (we formally here consider the full propagator expression for the
$\chi$ field. This will be convenient later, and we use the
approximation that the spectral function for $\chi$ has a
Breit-Wigner form, with width $\Gamma_\chi$, to write the
expressions for the propagator as \cite{BR})

\begin{eqnarray}
&& G^> ({\bf q}, t-t') = \frac{1}{2 \omega_\chi} \left\{
\exp[-i (\omega_\chi - i \Gamma_\chi) (t-t')] \theta(t-t') +
\exp[-i (\omega_\chi + i \Gamma_\chi) (t-t')] \theta(t'-t) \right\}
\nonumber \\
&& G^< ({\bf q}, t-t') = G^> ({\bf q}, t'-t) \;.
\label{G><}
\end{eqnarray}
{}For the analogous expressions at finite temperature, see for
example Ref. \cite{BR}.

In terms of the field variables $\phi_+$ and $\phi_-$ (in the forward
and backward branches of the time contour, respectively) and the
real time propagator Eq. (\ref{Gmatrix}), the effective action to
order  $g^4$ becomes

\begin{eqnarray}
\Gamma[\phi_+,\phi_-] &=& \int d^4 x \left\{ \left[
\frac{1}{2} (\partial_\mu \phi_+)^2 - \frac{m_\phi^2}{2} \phi_+^2
\right]
- \left[ \frac{1}{2} (\partial_\mu \phi_-)^2 - \frac{m_\phi^2}{2}
\phi_-^2 \right] \right\} \nonumber \\
&-& \frac{g^2}{2} \int d^4 x \int \frac{d^3 q}{(2 \pi)^3}
\left[ \phi_+ (x) G_\chi^{++}({\bf q},0) - \phi_-(x) G_\chi^{--}({\bf
q},0)
\right] \nonumber \\
&+& i \frac{g^4}{4} \int d^4 x d^4x' \int \frac{d^3 k}{(2 \pi)^3}
\exp[i {\bf k}.({\bf x}-{\bf x}')] \int \frac{d^3 q}{(2 \pi)^3}
\left[ \phi_+(x) G_\chi^{++} ({\bf k}+ {\bf q}, t-t')
G_\chi^{++} ({\bf q}, t-t') \phi_+(x') \right. \nonumber \\
&-& \left.
\phi_+(x) G_\chi^{+-} ({\bf k}+ {\bf q}, t-t')
G_\chi^{+-} ({\bf q}, t-t') \phi_-(x')
- \phi_-(x) G_\chi^{-+} ({\bf k}+ {\bf q}, t-t')
G_\chi^{-+} ({\bf q}, t-t') \phi_+(x') \right. \nonumber \\
&+& \left.
\phi_-(x) G_\chi^{--} ({\bf k}+ {\bf q}, t-t')
G_\chi^{--} ({\bf q}, t-t') \phi_-(x') \right]\;.
\label{Gamma+-}
\end{eqnarray}
At this point it is more convenient to introduce two new variables
$\phi_c$ and $\phi_\Delta$, given in terms of $\phi_+$ and $\phi_-$
by

\begin{eqnarray}
&& \phi_c =\frac{\phi_++\phi_-}{2}\;,\nonumber \\
&& \phi_\Delta = \phi_+ - \phi_- \;.
\label{new phi}
\end{eqnarray}
Rewriting the effective action (\ref{Gamma+-}) in terms of
$\phi_c$ and $\phi_\Delta$, we obtain

\begin{eqnarray}
\Gamma[\phi_c,\phi_\Delta] &=& \int d^4 x \phi_\Delta (x) [-\Box -
m_\phi^2]
\phi_c (x) - \frac{g^2}{2} \int d^4 x \phi_\Delta (x)
\int \frac{d^3 q}{(2 \pi)^3} \frac{1}{2 \omega_\chi} \nonumber \\
&+& i \frac{g^4}{4} \int d^4 x d^4 x'  \phi_\Delta (x)\phi_\Delta(x')
\int \frac{d^3 k}{(2 \pi)^3}
\exp[i {\bf k}.({\bf x}-{\bf x}')] \int \frac{d^3 q}{(2 \pi)^3}
{\rm Re} [G_\chi^{++} ({\bf k} + {\bf q}, t-t')
G_\chi^{++} ({\bf q}, t-t')] \nonumber \\
&-& g^4 \int d^4 x d^4 x'  \phi_\Delta (x)\phi_c (x')
\int \frac{d^3 k}{(2 \pi)^3}
\exp[i {\bf k}.({\bf x}-{\bf x}')] \nonumber \\
&\times&\int \frac{d^3 q}{(2 \pi)^3}
{\rm Im} [G_\chi^{++} ({\bf k} + {\bf q}, t-t')
G_\chi^{++} ({\bf q}, t-t')] \theta(t-t')\;.
\label{Gammanew}
\end{eqnarray}
The imaginary term in Eq. (\ref{Gammanew}) can be associated as
coming from a functional integration over a Gaussian noise
(stochastic) field $\xi_\phi$ \cite{GR}

\begin{eqnarray}
&&\int D\xi_\phi P[\xi_\phi] \exp\left\{i \int
d^4 x \phi_{\Delta} (x) \xi_\phi (x) \right\} \nonumber \\
&=& \exp\left\{- \int d^4 x d^4 x' \frac{g^4}{4}
\phi_{\Delta} (x) {\rm Re}\left[G_\chi^{++}G_\chi^{++}\right]_{x,x'}
\phi_{\Delta} (x') \right\}\: ,
\label{noise}
\end{eqnarray}
where $P[\xi_\phi]$ is the probability distribution for
$\xi_\phi$ and it is given by

\begin{equation}
P[\xi_\phi] = N^{-1} \exp\left\{- \frac{1}{2} \int d^4 x d^4 x'
\xi_\phi (x)
\left( \frac{g^4}{2} {\rm Re} \left[
 G_\chi^{++}G_\chi^{++}\right]_{x,x'}
\right)^{-1} \xi_\phi (x') \right\} \: .
\label{Pxi}
\end{equation}
In Eqs. (\ref{noise}) and (\ref{Pxi}) we have used the short notation
for the product of propagators:

\begin{equation}
\left[G_\chi^{++} G_\chi^{++}\right]_{x,x'} = \int \frac{d^3 k}{(2
 \pi)^3}
\exp[i {\bf k}.({\bf x}-{\bf x}')] \int \frac{d^3 q}{(2 \pi)^3}
[G_\chi^{++} ({\bf k} + {\bf q}, t-t')
G_\chi^{++} ({\bf q}, t-t')] \;.
\label{GG}
\end{equation}

Using Eq. (\ref{noise}) then we can write Eq. (\ref{Gammanew}) as

\begin{equation}
\Gamma[\phi_c, \phi_\Delta] = -i \ln \int D\xi_\phi P[\xi_\phi]
\exp \left\{ i S_{\rm eff}[\phi_c,\phi_\Delta,\xi_\phi]\right\}\;,
\label{action2}
\end{equation}
where

\begin{eqnarray}
S_{\rm eff} [\phi_c, \phi_\Delta, \xi_\phi] &=&
\int d^4x \phi_\Delta(x)[-\Box - m_\phi^2] \phi_c(x) -
\sum_{j=1}^{N_\chi}
\frac{g^2}{2} \int d^4x \phi_\Delta (x) \int \frac{d^3 q}{(2 \pi)^3}
\frac{1}{2 \omega_\chi} \nonumber \\
&-& \sum_{j=1}^{N_\chi} g^4 \int d^4x d^4x' \phi_\Delta (x)
\phi_c(x')
{\rm Im}\left[G_\chi^{++} G_\chi^{++}\right]_{x,x'} \theta(t-t')
+ \int d^4 x \phi_\Delta (x) \xi_\phi(x) \;.
\label{Seff}
\end{eqnarray}
The second term in the above equation is divergent. It can be easily
removed  by adding the appropriate renormalization counter-term in
the classical potential. In the following we assume a renormalized
action and just drop that divergent term from the equations.

The equation of motion for $\phi_c$ is defined by (see for instance
Ref.~\cite{GR})

\begin{equation}
\frac{\delta S_{\rm eff}[\phi_{\Delta},\phi_c,\xi_\phi]}
{\delta \phi_{\Delta}}|_{\phi_{\Delta} = 0} = 0 ~~,
\label{motion}
\end{equation}
from which we obtain

\begin{equation}
(\Box + m_\phi^2) \phi_c (x) + \sum_{j=1}^{N_\chi} g^4 \int d^4x'
\phi_c (x')
{\rm Im}\left[G_\chi^{++} G_\chi^{++}\right]_{x,x'} \theta(t-t')
= \xi_\phi (x)\;.
\label{eomphic}
\end{equation}
Note, from Eqs. (\ref{noise}) and (\ref{Pxi}), that the noise field
$\xi_\phi$ satisfies

\begin{eqnarray}
&& \langle \xi_\phi \rangle =0 \nonumber \\
&& \langle \xi_\phi (x) \xi_\phi (x') \rangle = \sum_{j=1}^{N_\chi}
\frac{g^4_j}{2}
{\rm Re}\left[G_{\chi_j}^{++} G_{\chi_j}^{++}\right]_{x,x'}\;.
\label{noise prop}
\end{eqnarray}

Taking the (ensemble) average of Eq. (\ref{eomphic}) and associating
$\langle \phi_c \rangle $ to the classical inflaton field
configuration, we obtain (for a homogeneous field, $\varphi \equiv
\varphi(t)$)

\begin{equation}
\ddot{\varphi}(t) + m_\phi^2 \varphi(t) +
\sum_{j=1}^{N_\chi} g_j^4 \int_{-\infty}^t d t' \varphi(t')
\int \frac{d^3q}{(2 \pi)^3}
{\rm Im}[G_{\chi_j}^{++}({\bf q}, t-t')]_{t>t'}^2 =0 \;.
\label{eomphi2}
\end{equation}
This equation is exactly the same as the one obtained from the
tadpole method, Eq. (\ref{eom2}).

Using Eqs. (\ref{G of q}) and (\ref{G><}), we obtain that

\begin{equation}
{\rm Im}[G_{\chi_j}^{++}({\bf q}, t-t')]_{t>t'}^2 =
-\frac{\exp(-2 \Gamma_{\chi_j} |t-t'|)}{4 \omega_{\chi_j}^2}
\sin(2 \omega_{\chi_j} |t-t'|)\;.
\label{ImGG}
\end{equation}
Substituting this into Eq. (\ref{eomphi2}), we obtain

\begin{equation}
\ddot{\varphi}(t) + m_\phi^2 \varphi(t) -
\sum_{j=1}^{N_\chi} g_j^4 \int_{-\infty}^t d t' \varphi(t')
\int \frac{d^3q}{(2 \pi)^3}
\frac{\exp(-2 \Gamma_{\chi_j} |t-t'|)}{4 \omega_{\chi_j}^2}
\sin(2 \omega_{\chi_j} |t-t'|)=0 \;.
\label{eomphi3}
\end{equation}
By integrating by parts in the time the last term in the above, we
obtain

\begin{equation}
\ddot{\varphi}(t) + \bar{m}_\phi^2 \varphi(t) +
\sum_{j=1}^{N_\chi} g_j^4 \int_{-\infty}^t d t' \dot{\varphi}(t')
\int \frac{d^3q}{(2 \pi)^3}
\frac{\exp(-2 \Gamma_{\chi_j} |t-t'|) \left[
\omega_{\chi_j} \cos(2 \omega_{\chi_j} |t-t'|)+
\Gamma_{\chi_j} \sin(2 \omega_{\chi_j} |t-t'|)
\right]}{8 \omega_{\chi_j}^2
(\Gamma_{\chi_j}^2 +\omega_{\chi_j}^2)} =0 \;,
\label{eomphi4}
\end{equation}
where

\begin{equation}
\bar{m}_\phi^2 = m_\phi^2 - \sum_{j=1}^{N_\chi} g_j^4 \int
\frac{d^3q}{(2\pi)^3}
\frac{1}{8 \omega_{\chi_j} (\Gamma_{\chi_j}^2 +
\omega_{\chi_j}^2)}\;.
\label{mphi}
\end{equation}
Using now that the $\chi_j$ fields are allowed to decay into the
$\psi_k$ fields with a decay rate $\Gamma_{\chi_j}(q)$ given by
\cite{BR}

\begin{equation}
\Gamma_{\chi_j} (q) = \sum_{k=1}^{N_\psi} \frac{h_k^2
m_{\chi_j}^2}{8 \pi \omega_{\chi_j}({\bf q})}
\left(1 - 4 \frac{m_{\psi_k}^2}{m_{\chi_k}^2} \right)^{3/2} =
\frac{\alpha_{\chi_j}^2}{\omega_{\chi_j}({\bf q})}\;,
\label{Gammachi}
\end{equation}
where $\alpha_{\chi_j}^2= \sum_{k=1}^{N_\psi}h_k^2
m_{\chi_j}^2 (1 - 4 m_{\psi_k}^2/m_{\chi_k}^2)^{3/2}/(8 \pi)$.
Substituting this into Eq. (\ref{eomphi4}) and approximating the
dissipative kernel by a Markovian one (this is a valid approximation
provided we have a sufficiently large number of decay channels
available, see \cite{BR}) we find the local equation of motion for
$\varphi$

\begin{equation}
\ddot{\varphi}(t) + \bar{m}_\phi^2 \varphi(t) +
\sum_{j=1}^{N_\chi} g_j^4 \dot{\varphi}(t) \int
\frac{d^3q}{(2\pi)^3}
\frac{\Gamma_{\chi_j}(q)}{8 \omega_{\chi_j} (\omega_{\chi_j}^2+
\Gamma_{\chi_j}^2)^2} =0\;.
\label{eom markov}
\end{equation}
Substituting  Eq. (\ref{Gammachi}) in the above equation,
the momentum integral in Eq. (\ref{eom markov}) can be performed and
the final result for the effective EOM for the homogeneous
field $\varphi$:

\begin{equation}
\ddot{\varphi}(t) + \bar{m}_\phi^2 \varphi(t) + \eta_\phi \dot{\varphi}(t)=0\;,
\label{final eom}
\end{equation}
where

\begin{equation}
\eta_\phi = \sum_{j=1}^{N_\chi} \frac{g_j^4 \alpha_{\chi_j}^2}{128 \pi
(m_{\chi_j}^4+
\alpha_{\chi_j}^4)^{\frac{1}{2}} (2 \sqrt{m_{\chi_j}^4+
\alpha_{\chi_j}^4} + 2 m_{\chi_j}^2)^{\frac{1}{2}}}\;.
\label{etaphi}
\end{equation}
This equation has a simple damped solution for $\varphi(t)$
(overdamped for $\eta_\phi > 2 \bar{m}_\phi$, underdamped
for $\eta_\phi < 2 \bar{m}_\phi$).

\subsection{The Equation of Motion for the Fluctuations $\chi_j$}

We can derive the effective dissipative and stochastic equation for the
fluctuation fields $\chi$ in a similar way. The perturbative effective
action for $\chi$ for the model (\ref{model}) is \footnote{In the
following we neglect the corrections of the fermions to $\chi_j$, given
by $\langle \bar{\psi_k} \psi_k \rangle$. This is fine in the Markovian
limit, in which case the contributions from the fermions to dissipation
and noise vanish as a consequence of having $\Gamma_{\psi_k}=0$
\cite{BR} and the fermions only give a renormalization to the $\chi_j$
fields mass.} (in analogy with Eq. (\ref{Gammanew}))

\begin{eqnarray}
\Gamma[\chi_c,\chi_\Delta] &=& \int d^4 x \chi_\Delta (x) [-\Box - m_\chi^2
-g^2 \varphi]
\chi_c (x)\nonumber \\
&+& i \frac{g^4}{2} \int d^4 x d^4 x'  \chi_\Delta (x)\chi_\Delta (x')
{\rm Re} [G_\chi^{++} G_\phi^{++}]_{x,x'} \nonumber \\
&-& 2 g^4 \int d^4 x d^4 x'  \chi_\Delta (x)\chi_c (x')
{\rm Im} [G_\chi^{++} G_\phi^{++}]_{x,x'} \theta(t-t')\;.
\label{actionchi}
\end{eqnarray}
Again, we can interpret the imaginary term in the above equation as coming
from a functional integration over a noise field $\xi_\chi$:

\begin{eqnarray}
&&\int D\xi_\chi P[\xi_\chi] \exp\left\{i \int
d^4 x \chi_{\Delta} (x) \xi_\chi (x) \right\} \nonumber \\
&=& \exp\left\{- \int d^4 x d^4 x' \frac{g^4}{2}
\chi_{\Delta} (x) {\rm Re}\left[G_\chi^{++}G_\phi^{++}\right]_{x,x'}
\chi_{\Delta} (x') \right\}\: ,
\label{noisechi}
\end{eqnarray}
where $P[\xi_\chi]$ is the probability distribution for
$\xi_\chi$ and it is given by

\begin{equation}
P[\xi_\chi] = N^{-1} \exp\left\{- \frac{1}{2} \int d^4 x d^4 x' \xi_\chi (x)
\left( g^4 {\rm Re} \left[ G_\chi^{++}G_\phi^{++}\right]_{x,x'}
\right)^{-1} \xi_\chi (x') \right\} \: .
\label{Pxichi}
\end{equation}
Rewriting the action for $\chi$ in terms of the noise field $\xi_\chi$, then
the equation of motion which we obtain (analogous to Eq. (\ref{eomphic}))
is

\begin{equation}
(\Box + m_\chi^2 + g^2 \varphi ) \chi_c (x) + 2 g^4 \int d^4x' \chi_c (x')
{\rm Im}\left[G_\chi^{++} G_\phi^{++}\right]_{x,x'} \theta(t-t')
= \xi_\chi (x)\;.
\label{eomchic}
\end{equation}
In order to get an approximate analytical expression for the above equation
of motion, we will assume only contributions to the effective action
with zero external momentum (the linear response approximation) or
nearly spatially homogeneous fields. This way we can handle the spatial
nonlocality in Eq. (\ref{eomchic}) and obtain:

\begin{equation}
(\Box + m_\chi^2 + g^2 \varphi ) \chi_c (x) - 2 g^4 \int_{-\infty}^t dt'
\chi_c ({\bf x},t')
\int  \frac{d^3q}{(2 \pi)^3}
\frac{\exp[-(\Gamma_\chi+\Gamma_\phi) |t-t'|]}{4 \omega_\chi \omega_\phi}
\sin[(\omega_\chi+\omega_\phi) |t-t'|]
= \xi_\chi (x)\;.
\label{eomchic2}
\end{equation}
Integrating by parts in the time the last term in the lhs of the
above equation, we obtain

\begin{eqnarray}
&&(\Box + \bar{m}_\chi^2 + g^2 \varphi ) \chi_c (x)
+ 2 g^4 \int_{-\infty}^t dt'
\dot{\chi}_c ({\bf x},t')
\int  \frac{d^3q}{(2 \pi)^3} \nonumber \\
&& \times
\frac{\exp[-(\Gamma_\chi+\Gamma_\phi) |t-t'|]
\left\{(\Gamma_\chi+\Gamma_\phi)
\sin[(\omega_\chi+\omega_\phi) |t-t'|]
+(\omega_\chi+\omega_\phi)\cos[(\omega_\chi+\omega_\phi) |t-t'|]
\right\}}{4 \omega_\chi \omega_\phi
\left[(\Gamma_\chi+\Gamma_\phi)^2+(\omega_\chi+\omega_\phi)^2\right]}
= \xi_\chi (x)\;.
\label{eomchic3}
\end{eqnarray}
Once again, supposing valid a Markovian approximation for the dissipative
kernel in the above equation, we obtain for the effective EOM for the
fluctuation field $\chi_j$ the expression

\begin{equation}
(\Box + \bar{m}_{\chi_j}^2 + g_j^2 \varphi ) \chi_j (x) + \eta_{\chi_j}
\dot{\chi}_j(x) = \xi_{\chi_j}(x) \;,
\label{final eomchi}
\end{equation}
where

\begin{equation}
\eta_{\chi_j} = g_j^4 \int  \frac{d^3q}{(2 \pi)^3}
\frac{(\omega_{\chi_j} + \omega_\phi)(\Gamma_{\chi_j} + \Gamma_\phi)}{
\omega_{\chi_j} \omega_\phi \left[ (\omega_{\chi_j}+\omega_\phi)^2 +
(\Gamma_{\chi_j} +\Gamma_\phi)^2\right]^2}\;,
\label{etachi}
\end{equation}
while the effective (unrenormalized) mass $\bar{m}_{\chi_j}^2$ appearing
in Eq. (\ref{final eomchi}) and adding the contribution coming from
the fermions, we find that

\begin{equation}
\bar{m}_{\chi_j}^2 = m_{\chi_j}^2 - \frac{g_j^4}{2} 
\int  \frac{d^3q}{(2 \pi)^3}
\frac{\omega_{\chi_j} + \omega_\phi}{
\omega_{\chi_j} \omega_\phi \left[ (\omega_{\chi_j}+\omega_\phi)^2 +
(\Gamma_{\chi_j} +\Gamma_\phi)^2\right]} +
2 \sum_{k=1}^{N_\psi} h_k^2
\int\frac{d^3 q}{(2 \pi)^3} \frac{m_{\psi,r}^2-\omega_{\psi_k}^2}
{\omega_{\psi_k}^3}
\label{mchi}
\end{equation}

{}For the model (\ref{model}) we have $\Gamma_{\chi_j}$ as given by
Eq. (\ref{Gammachi}) and $\Gamma_\phi$ is
given by

\begin{equation}
\Gamma_\phi (q) = \sum_{j=1}^{N_\chi} 
\frac{g_j^4}{16 \pi \omega_\phi} \left(1 - 4
\frac{m_{\chi_j}^2}{m_\phi^2}\right)^{1/2} \;.
\label{Gammaphi}
\end{equation}
The noise field in Eq. (\ref{final eomchi}), from Eqs. (\ref{noisechi}) and
(\ref{Pxichi}), obeys

\begin{eqnarray}
&&\langle \xi_\chi \rangle =0 \;,\nonumber \\
&&\langle \xi_\chi (x) \xi_\chi(x')\rangle = g_j^4 {\rm Re}
[G_{\chi_j}^{++} G_{\phi_j}^{++}]_{x,x'} \;.
\end{eqnarray}

\section{The (In)stability Bands and their borders}

The equations (\ref{final eom}) and (\ref{final eomchi}), obtained in
the previous sections, are the basic equations of this paper. If one
neglects their dissipation and inhomogeneous terms, the system will
present the exact parametric resonance phenomenon, with instability
bands in the parameter space inside which modes grow indefinitely, with
a real (and positive) Floquet exponent. Here, we make no such
approximation. We only restrict ourselves to a time period when the
back-reaction can be safely neglected, {\it i.e.}, in the beginning of the
preheating phase.

Our goal is twofold. First, we check the very existence of the band
structure when one uses the set of equations mentioned above, obtained
from a microscopic calculation, as opposed to inserting {\it ad hoc}
terms in their classical counterparts. As we will shortly see, the
parametric resonance is barely changed by this modification. Second, we
investigate the dimension of the borders of the bands. It has been known
\cite{levin} that a fractal border prevents one from determining the
final attractor for which a given trajectory will tend to. Although we
cannot formally define an attractor in our problem, we can split the
trajectories in two different cases: the one which are amplified and the
ones which are not. We will explain the exact procedure used later on
this section.

The instability bands are manifest in the momentum space $k$ for the Eq.
(\ref{final eomchi}). Actually, we will plot the bands in the space $k^2
\times g^2$, as has become usual in the literature. In momentum space
Eq. (\ref{final eomchi}) becomes

\begin{equation}
\ddot{\chi}_j ({\bf k},t) + \left[{\bf k}^2 + 
\bar{m}_{\chi_j}^2 + g_j^2 \varphi(t)
\right] \chi_j ({\bf k},t) + \eta_{\chi_j}
\dot{\chi}_j( {\bf k},t) = \xi_{\chi_j}({\bf k},t) \;,
\label{eomchik}
\end{equation}
which together with Eq. (\ref{final eom}) form a coupled system of
differential equations that we study numerically. The typical values of
parameters that we use are \cite{brand}

\begin{eqnarray}
g_j &\equiv& g =  10^{-4} M_{Pl}^{1/2}\nonumber\\
h_k &\equiv& h = 0.1 \nonumber\\
\varphi(t=0)&=& M_{Pl}/5 \nonumber\\
\chi_j({\bf k},t=0) &\approx& 1/M_{Pl}^2  \\
m_\phi &=& 10^{-6} M_{Pl} \nonumber\\
m_\chi &=& 3/7 \times 10^{-6} M_{Pl} \nonumber\\
m_\psi &=& 1/7 \times 10^{-6} M_{Pl} \nonumber
\end{eqnarray}
where we have expressed all dimensionfull quantities
with respect to the Planck mass
$M_{Pl}$.

\begin{figure}[ht]
\epsfysize=15cm
{\centerline{\epsfbox{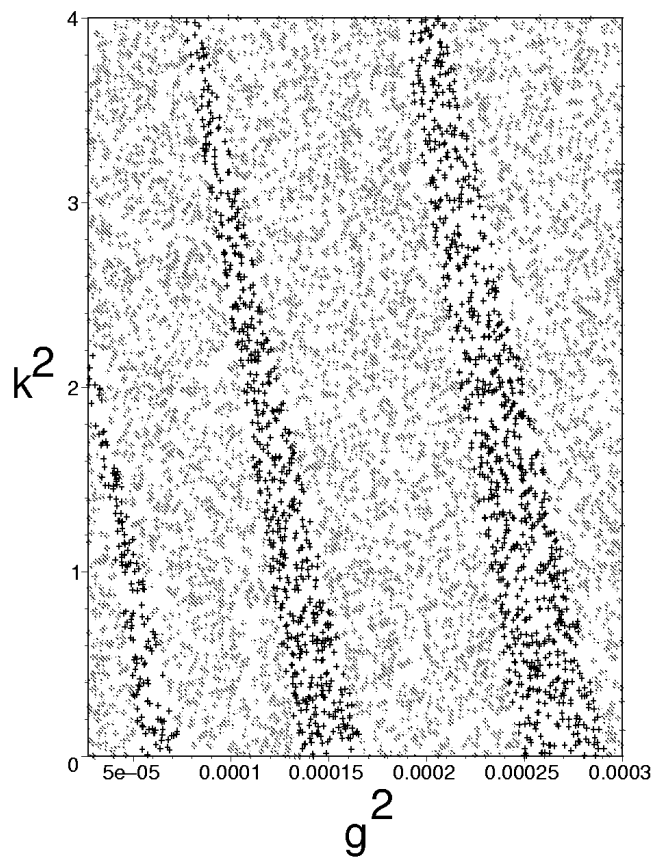}}}
\caption{Band structure in the $k^2 \times g^2$ space. Dark crosses indicate
final energy above the cutoff (which, in this case, equals $100\,E_0$).}
\label{bandfig}
\end{figure}

The chaotic behavior of the dynamical system of equations, as
determined by the equations of motion for $\varphi(t)$,
Eq.~(\ref{final eom}), and the one for the fluctuation field modes,
$\chi_j({\bf k}, t)$, Eq. (\ref{eomchik}), is then quantified  by
means of the determination of the fractal dimension of the boundary
separating each of the instabilities bands. To define exactly what is
to be called an {\it unstable} band we establish a set of cutoff
values for the energy $E$ of the $\chi$-field: $E_0$, $10 E_0$ and
$100 E_0$, where $E_0$ is the largest possible initial energy (in momentum
space) in the
range of initial conditions taken,

\begin{equation}
E_0 =
\frac{1}{2} \left[ \dot{\chi_j}^2 ({\bf k},t_0) +k^2 \chi_j^2({\bf k},t_0) +
m_{\chi_j}^2 \chi_j^2({\bf k},t_0)
+g_j^2 \varphi (t_0) \chi_j^2 ({\bf k},t_0)\right] \;.
\end{equation}
{}Fig.~\ref{bandfig} shows the band structure in the phase space and the
set of first three instability bands we will restrict ourselves to. Each
energy cutoff defines the border between two different ranges for the Floquet
exponent.

We next move to the determination of the fractal dimension.
The fractal dimension is associated with the possible different exit
modes under small changes of the initial conditions at $t=0$
and it will give a measure of the degree of chaos of our dynamical
system. The exit modes we refer to above are the ones associated to
trajectories that evolve to the stable band or to the unstable one,
as determined by the given cutoff energy $E_c$. The method we employ
to determine the fractal dimension is the box-counting method, which
is a standard method for determining the fractal dimension of
boundaries \cite{ott}. Its definition and the specific numerical
implementation we use here have been described in details in Ref.
\cite{fractalpaper}.

The basic procedure is: given a set of initial conditions ${\bf x}_0$
at $t=t_0$ that leads to a certain outcome for the trajectory in
phase space, we study whether there will be a change of outcome
for the trajectory due to a perturbation $\delta$ (i.e., whether
the perturbation will lead to a different attractor or not). Given a
volume region in phase space around a boundary between different
attractors and perturbing a large set of initial conditions inside
that region, the fraction of uncertain trajectories, $f(\delta)$,
that result in a different outcome under a small perturbation can be
shown to scale with the perturbation $\delta$ as \cite{ott}
$f(\delta) \sim \delta^\epsilon$, where $\epsilon$ is called the
uncertainty exponent. The box-counting dimension of the boundary in
phase space separating different attractors, or fractal dimension
$f_d$, is given by~\cite{ott} $f_d = d -\epsilon$, where $d$ denotes
the dimension of the phase space. For our system of equations of
motion, $d=4$. For a fractal boundary $f_d > d -1$, implying that
$\epsilon < 1$, whereas for a non-fractal boundary, $f_d = d - 1$,
and $\epsilon = 1$.

{}Following the method of box-counting for this problem, it is enough
to consider variations only on the parameters $k^2$ and $g^2$, the
wave-number squared and coupling constant squared, respectively.
Perturbations around these variables are taken between $10^{-5}$ to
$10^{-4}$, as we move from the first to the third band in Fig. 1 and
where the initial conditions are determined. For each region
separating the instability bands, set by the variations of $k^2$ and
$g^2$, we then consider a large number of random points (a total of
$10.000$ random points for each run was used) enough to produce a
reliable statistics with errors around the percent (see
Ref.~\cite{fractalpaper} for a description of estimates of the
statistical error). All initial conditions are then numerically
evolved by using an eighth-order Runge-Kutta integration method and
the fractal dimension is obtained by statistically studying the
outcome of each initial condition for each run of the large set of
points.

The fractal dimension of each of the borders shown in Fig. 1
is given in Table I. Note that they all differ only within the
statistical error in the determination of the fractal dimension.
We have verified that this result also applies to other different
regions of parameter space, showing that this is not a coincidence
for the instability bands shown. This is an important result since
for every parameter region chosen, they will all have similar
dynamical properties with each instability band then sharing similar
properties. In this case, the dynamics is much less affected by the
parameter space but only model dependent. In special, the fractality
of those borders  shows that one cannot be certain about the precise
value of the Floquet exponent for a given mode, even though it is
well inside an instability band. The numerical results are shown in 
Table \ref{bands}, for the first three bands and for three different 
energy cutoff values. The results shown were produced considering
$N_\chi=N_\psi=1$ in Eqs. (\ref{final eom}) and (\ref{final eomchi}).

\begin{figure}[ht]
\epsfxsize=12cm
\epsfysize=12cm
{\centerline{\epsfbox{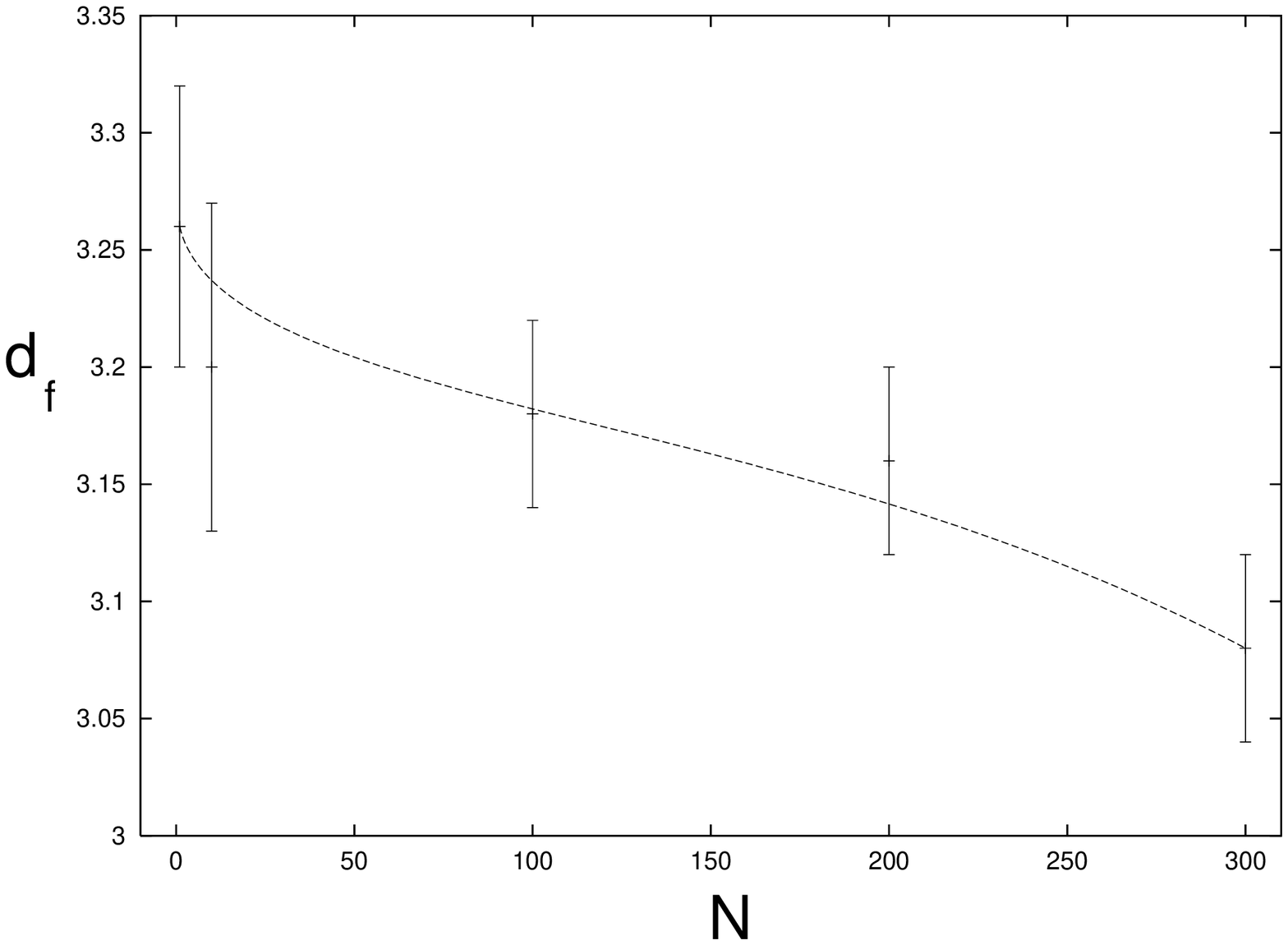}}}
\caption{Fractal dimension $d_f$ of the first band 
shown in {}Fig. 1 {\it versus} the
number of interacting fields $N$. }
\label{ndf}
\end{figure}

We next study the influence of the number of field modes coupled to
the inflaton field $\Phi$, by changing the number of fermion fields $N_\psi$
and the boson fields ones $N_\chi$. As it can be seen from Eqs. (\ref{etaphi})
and (\ref{etachi}) this will, consequently, lead to changes to the
dissipation coefficients. As we increase the decaying modes to both
$\Phi$ and $\chi_j$ fields we expect the larger be the dissipation
terms. The damping of the oscillations of the background field will
eventually lead to a destruction of the instability bands. At the same time,
even for underdamping oscillations we expect that as we increase the decaying
modes the smoother will get the borders between stable and unstable
regions of parameter space, leading to a continuous decrease of the
fractal dimension associated to the borders of the bands. This is
confirmed by our numerical studies and the results for the change in
fractal dimension as we increase the number of fields is shown
in Fig. \ref{ndf}, where we have restricted ourselves to the first
instability band shown in {}Fig. 1, for an energy threshold of $E= 100
E_0$, and considered for simplicity $N_\chi=N_\psi=N$. 
Similar results would also follow to the other bands but are
not shown here.

\section{Conclusions}

Our results indicate that the fractal dimension of the borders (within
the statistical error) does not change in any significant way for each
instability band. It also shows very little dependence on the energy
threshold chosen. These results, besides indicating that the border
between the regions (defined by their own Floquet exponents) inside a
given band is itself fractal and thus exhibits chaotic dynamics, shows
that the dynamical properties of each instability band is independent of
the precise region taken in the parameter space, and it is only model
dependent.

The fractal structure is also present if we fix the parameters $k$ and
$g$ and vary only the initial inflaton conditions ($\varphi$ and
$\dot{\varphi}$). This shows a strong dependence of the outcome on the
initial conditions, with major implications in the early universe
models. {}For instance, an initial phase of warm inflation
\cite{BR,BGR2}, even if it is not able to generate enough radiation to
match the standard Big Bang model, may turn out to be crucial for
generating the adequate initial conditions for the preheating phase.

One may notice from Eq. (\ref{eomchik}) that the effective mass squared
$m_{\rm{eff}}^2\equiv {\bf k}^2 + \bar{m}_{\chi_j}^2 + g_j^2 \varphi(t)$
for the field $\chi_k$ may become negative, triggering its decay via the
so-called tachyonic instability (also known as spinodal decomposition).
We have restricted the initial value of the inflaton $\phi$ so that it does not
happen in our simulations. Nevertheless, the combination of such a decay
with parametric amplification is an interesting and complex subject on
its own and it will be addressed in a forthcoming publication
\cite{future}.

\acknowledgments

The authors would like to thank R. Brandenberger for many discussions in
the subject of this paper. R.O.R. and S.E.J. were supported by Conselho
Nacional de Desenvolvimento Cient\'{\i}fico e Tecnol\'ogico - CNPq
(Brazil). S.E.J. acknowledges partial support from the U.S. Department
of Energy under the contract DE-FG02-91ER40688 - Task A, at Brown
University. R.O.R. was also supported by the ``Mr. Tompkins Fund for
Cosmology and Field Theory'' at Dartmouth College, where this work 
has started.

\newpage

\begin{table}[t]
\vspace{1cm}
\begin{center}
 \begin{tabular}{c|c|c|c}
   \hline\hline
   $E$       &  $d_f^I$        & $d_f^{II}$        & $d_f^{III}$ \\
   \hline
   $E_0$     &  $3.26\pm 0.06$  & $3.28 \pm 0.09$ & $3.22 \pm 0.07$  \\
   \hline
   $10 E_0$  &  $3.23\pm 0.08$  &  $3.23 \pm 0.09$ &  $3.24 \pm 0.05$ \\
   \hline
   $100 E_0$ &  $3.20\pm 0.05$  &  $3.20 \pm 0.07$ &  $3.16 \pm 0.06$ \\
   \hline\hline
 \end{tabular}
\end{center}
\caption{Fractal dimensions $d_f$ for the first (I), second (II) and
third (III) instability bands, at different energy thresholds (E).
Bands are numbered according to increasing $g^2$, {\it i.e.}, 
from left to right in {}Fig. \ref{bandfig}.}
\label{bands}
\end{table}

\vspace{1cm}

\begin{thebibliography}{99}

\bibitem{reheat}L. Kofman, A. Linde and A. A. Starobinsky,
Phys. Rev. Lett. {\bf 73}, 3195 (1994); {\it ibid.},
{\bf 76},1011 (1996); Y. Shtanov, J. Traschen and R. Brandenberger,
Phys. Rev. {\bf D51}, 5438 (1995).


\bibitem{preheat}L. Kofman, A. Linde and A. A. Starobinsky,
Phys. Rev. {\bf D56}, 3258 (1997); D. Boyanovsky, H. J. de Vega,
R. Holman and J. F. J. Salgado, Phys. Rev. {\bf D54}, 7570 (1996).

\bibitem{joras}S. E. Jor\'as and V. H. Cardenas, gr-qc/0108088.

\bibitem{felder}G. Felder and L. Kofman, Phys. Rev. {\bf D63},
103503 (2001).

\bibitem{RF}R. O. Ramos and F. A. R. Navarro, Phys. Rev.
{\bf D62}, 085016 (2000). 

\bibitem{ROR}R. O. Ramos, Phys. Rev. {\bf D64}, 123510 (2001).

\bibitem{GR}M. Gleiser and R. O. Ramos, Phys. Rev. {\bf D50}, 2441 (1994).

\bibitem{noise1}V. Zanchin, A. Maia, Jr., W. Craig and
R. Brandenberger, Phys. Rev. {\bf D57}, 4651 (1998).

\bibitem{noise2} V. Zanchin, A. Maia, Jr., W. Craig and
R. Brandenberger, Phys. Rev. {\bf D60}, 023505 (1999). 

\bibitem{hu1}B. L. Hu, J. P. Paz and Y. Zhang, in {\it The Origin of
Structure in the Universe}, edited by E. Gunzig and P. Nardone (Kluwer
Academic, Norville, 1993); B. L. Hu, J. P. Paz and Y. Zhang, Phys.
Rev. {\bf D45}, 2843 (1993); {\it ibid} {\bf D47}, 1576 (1993); E.
Calzetta and B. L. Hu, Phys. Rev. {\bf D49}, 6636 (1994); {\it
ibid } {\bf D52}, 6770 (1995); B. L. Hu and A. Matacz, Phys. Rev.
{\bf D51}, 1577 (1995); A. Matacz, Phys. Rev. {\bf D55}, 1860
(1997).

\bibitem{morikawa}M. Morikawa, Phys. Rev. {\bf D33}, 3607 (1986).

\bibitem{hu2}S. A. Ramsey, B. L. Hu and A. M. Stylianopoulos,
Phys. Rev. {\bf D57}, 6003 (1998).

\bibitem{levin} J. D. Barrow and J. Levin, Phys. Rev. Lett.
{\bf 80},  656 (1998); N. J. Cornish and J. J. Levin, Phys.
Rev. {\bf D53}, 3022 (1996); Phys. Rev. {\bf D55}, 7489 (1997).

\bibitem{george} K. Ferraz, G. Francisco, G. E. A. Matsas, Phys.
Lett. {\bf  A156}, 407 (1991).

\bibitem{brand} Y. Shtanov, J. Traschen and R. Brandenberger, Phys. Rev {\bf
D51}, 5438 (1995); F. Finelli and R. Brandenberger, Phys. Rev {\bf D62}, 083502
(2000); Phys. Rev. Lett. {\bf 82}, 1362 (1999).

\bibitem{BR} A. Berera and R. O. Ramos, Phys. Rev. {\bf D63}, 103509 (2001).

\bibitem{BGR}A. Berera, M. Gleiser and R. O. Ramos, Phys. Rev. {\bf D58},
123508 (1998).

\bibitem{chou}K. Chou, Z. Su, B. Hao and L. Yu, Phys. Rep.
{\bf 118}, 1 (1985).

\bibitem{weert}N. P. Landsman and Ch. G. van Weert,  Phys. Rep.
{\bf 145}, 141 (1987).

\bibitem{rivers}R. Rivers, {\it Path Integral Methods in Quantum Field Theory}
(Cambridge University Press, Cambridge 1987).

\bibitem{ott} E. Ott, {\sl Chaos in Dynamical Systems} (Cambridge
University Press, Cambridge 1993).

\bibitem{fractalpaper}L. G. S. Duarte, L. A. C. P. da Mota, H. P. de Oliveira,
R. O. Ramos and J. E. Skea,
Comp. Phys. Comm. {\bf 119}, 256 (1999).

\bibitem{BGR2} A. Berera, M. Gleiser and R. O. Ramos, Phys. Rev. Lett.
{\bf 83}, 264 (1999). 

\bibitem{future} S. E. Jor\'as and R. O. Ramos, in preparation.

\end{thebibliography}
\end{document}